\documentclass[a4paper,11pt,reqno]{amsart}
\usepackage{amsmath,amsfonts,amssymb,enumerate}

\newtheorem{theorem}{Theorem}
\newtheorem{lemma}{Lemma}
\newtheorem{proposition}{Proposition}

\def\A{\mathcal{A}}

\def\R{\mathbb{R}}
\def\C{\mathbb{C}}

\def\Z{\mathbb{Z}}
\def\K{\mathbb{K}}
\def\g{\mathfrak{g}}

\def\p{\partial}
\def\proof{\noindent\textit{Proof. }}
\def\qed{$\blacksquare$}

\def\al{\alpha}
\def\CC{\mathbf{C}}

\def\KK{\mathbf{K}}

\def\brhd{\blacktriangleright}

\def\ad{\text{ad}}

\begin{document}

\title[Differential calculus on the Lie aglebra type noncommutative spaces]{Realization of bicovariant differential calculus on the Lie algebra type noncommutative spaces}

\author{Stjepan Meljanac}
\address{Ruder Bo\v{s}kovi\'{c} Institute, Theoretical Physics Division, Bijeni\v{c}ka c. 54, HR 10002 Zagreb, Croatia}
\email{meljanac@irb.hr}

\author{Sa\v{s}a Kre\v{s}i\'{c}--Juri\'{c}}
\address{Faculty of Science, Department of Mathematics, University of Split, Teslina 12, 21000 Split, Croatia}
\email{skresic@pmfst.hr}

\author{Tea Martini\'{c}}
\address{Faculty of Science, Department of Mathematics, University of Split, Teslina 12, 21000 Split, Croatia}
\email{teamar@pmfst.hr}

\date{}

\begin{abstract}
This paper investigates bicovariant differential calculus on noncommutative spaces of the Lie algebra type. For a given Lie algebra $\g_0$ we construct
a Lie superalgebra $\g=\g_0\oplus \g_1$ containing noncommutative coordinates and one--forms. We show that $\g$ can be extended by a set of generators $T_{AB}$
whose action on the enveloping algebra $U(\g)$ gives the commutation relations between monomials in $U(\g_0)$ and one--forms. Realizations of noncommutative
coordinates, one--forms and the generators $T_{AB}$ as formal power series in a semicompleted Weyl superalgebra are found. In the special case $\dim(\g_0)=
\dim(\g_1)$ we also find a realization of the exterior derivative on $U(\g_0)$. The realizations of these geometric objects yield a bicovariant differential
calculus on $U(\g_0)$ as a deformation of the standard calculus on the Euclidean space.
\end{abstract}


\maketitle


\section{Introduction}

This paper investigates construction of a differential calculus on quantum spaces using realizations of Lie superalgebras. One of the
greatest problems in modern physics is formulation of the theory of quantum gravity which consistently unifies Einstein's general
relativity with quantum mechanics. Einstein's theory of gravity coupled with Heisenberg's uncertainty principle implies that
space--time coordinates should satisfy uncertainty relations $\Delta x_\mu \Delta x_\nu \geq l_P^2$ where $l_P=\sqrt{G
\hbar/c^3} \approx 1.62\times 10^{-35}$ m is the Planck length \cite{Doplicher-1, Doplicher-2}. This requires a modification of
the usual notion of space--time as a continuum. One of the possible approaches to a description of space--time at the Planck
scale is based on replacing the commutative $C^\infty$ algebra of functions on a manifold by a noncommutative (NC) algebra of
operators $\hat x_1, \hat x_2,\ldots, \hat x_n$. This motivates the study of differential calculus on quantum spaces as a proper
mathematical tool for investigating physical theories in this context. Reformulation of the classical notions of differential
geometry in the algebraic language allows us to generalize them to the noncommutative setting \cite{Woronowicz,Schupp,Vladimirov}.
Noncommutativity between coordinates is introduced via commutation relations $[\hat x_\mu, \hat x_\nu]=\hat \theta_{\mu\nu}(h)$ where
$\hat \theta_{\mu\nu}(h)$ is an antisymmetric tensor depending on a deformation parameter $h\in \R$ and satisfying the classical
limit condition $\lim_{h\to 0} \hat \theta_{\mu\nu}(h)=0$. The algebraic relations between coordinates lead to different models
of NC spaces. The three most commonly studied NC spaces in the literature are the Moyal space \cite{Moyal, Groenewold},
$\kappa$--deformed space \cite{Lukierski, Zakrzewski, Majid} and Snyder space \cite{Snyder}. A review of applications of NC spaces
in physics can be found in Refs. \cite{Li, Aschieri}.

The general theory of differential calculus on quantum spaces was initiated and thoroughly investigated by Woronowicz in Ref. \cite{Woronowicz}. For a general associative
algebra, Landi gave a construction of a differential algebra of forms in Ref. \cite{Landi}. In analogy with classical differential geometry, Schupp \cite{Schupp} et. al.
introduced exterior derivative, inner derivations and Lie derivative on linear quantum groups. The requirement
that the differential calculus is bicovariant and also covariant under the imposed group of symmetries leads to some dimensionality problems. Sitarz \cite{Sitarz} has shown
that in order to construct a bicovariant calculus on $3+1$ dimensional $\kappa$--Minkowski space that is also Lorentz covariant, one has to
introduce an extra one--form that has no classical interpretation.  Gonera et. al. generalized this work to $n$ dimensions in Ref. \cite{Gonera}.
The differential complex and related geometric objects on the $\kappa$--Minkowski space were developed in Ref. \cite{Mercati}.
We note that for a given NC space the differential calculus is generally not unique. Classification of bicovariant differential calculi of classical
dimension on the $\kappa$--Minkowski space was given in Ref. \cite{Meljanac-1}.

In the present work we investigate a bicovariant first order differential calculus (FODC) on a general Lie algebra type NC space. Our approach uses
realizations of Lie superalgebras to construct geometric objects as deformations of the corresponding classical notions on the Euclidean space.
It generalizes our previous results on realizations of Lie algebras in Ref. \cite{SKJ} and some earlier attempts to construct differential
algebras on the $\kappa$--deformed space in Refs. \cite{Meljanac-1,Meljanac-2,Meljanac-3,Meljanac-4,Juric-1,Juric-2}.

The paper is organized as follows. In Section \ref{section-2} we outline briefly the construction of a bicovariant FODC of classical dimension
on the enveloping algebra $U(\g_0)$ where $\g_0$ is a finite dimensional Lie algebra. The generators of $U(\g_0)$ are interpreted as NC coordinates
satisfying a Lie algebra type commutation relations. Differential forms are introduced by extending $\g_0$ to a Lie superalgebra $\g=\g_0\oplus \g_1$
satisfying a compatibility condition required by exterior derivative. Section \ref{section-3} studies an extension of $\g$ by a set of generators
$T_{AB}$ which act on the enveloping algebra $U(\g)$. It is shown that the commutation relations between one--forms and monomials in $U(\g_0)$
can be expressed through the action of $T_{AB}$ on $U(\g)$. The Lie superalgebra $\g$ can be extended in two canonical ways related by
a left--right duality. In Section \ref{section-4} we construct realizations of $\g$ by representing the generators of $\g$ (NC coordinates and
one--forms) as formal power series in a semicompleted Weyl superalgebra (Clifford--Weyl algebra). Using an automorphism of the Weyl superalgebra we
find evidence that the Weyl symmetric realization of an arbitrary Lie algebra $\g_0$ (see Refs. \cite{SKJ} and \cite{Durov}) can be extended to Lie
superalgebras considered in this paper. Section \ref{section-5} deals with realization of the exterior derivative $d$ on $U(\g_0)$ in the special
case $\dim(\g_0)=\dim(\g_1)$. We construct $d$ as a deformation of the standard exterior derivative on $\R^n$ where the deformation parameters are
the structure constants of $\g$. The deformed exterior derivative $d$ is a nilpotent operator satisfying the undeformed Leibniz rule on $U(\g_0)$. Finally,
the Appendix provides details of some technical computations used in the paper.

\section{Bicovariant differential calculus on noncommutative spaces of the Lie algebra type}
\label{section-2}

Let $\g_0$ be a finite--dimensional Lie algebra over $\K$ ($\K=\R$ or $\C$). In this section we give a general construction of a bicovariant
first order differential calculus (FODC) on the enveloping algebra $U(\g_0)$. The construction uses extension of $\g_0$ to a Lie superalgebra
$\g=\g_0\oplus \g_1$ and it also motivates the study of realizations of $\g$ in later sections. First, let us describe the
Lie superalgebra $\g$. Let $C_{\mu\nu\lambda}$ denote the structure constants of $\g_0$ relative to a basis $X_1, X_2, \ldots, X_n$.
Let $\g_1$ be a vector space with basis $\theta_1, \theta_2, \ldots, \theta_m$.
Define the $\Z_2$--graded vector space $\g=\g_0\oplus \g_1$ with degrees of homogeneous elements given by $\bar X=i$ if $X\in \g_i$ for $i=0,1$.
For future use we adopt the following convention. The lowercase greek letters $\alpha, \beta, \gamma$, etc. range over the indices of $X_1, X_2,
\ldots, X_n$ and the lowercase latin letters $a,b,c$, etc. range over the indices of $\theta_1, \theta_2, \ldots, \theta_m$. If $n=m$, then the
indices of $\theta_1, \theta_2, \ldots, \theta_m$ will be denoted by lowercase greek letters.
Define the Lie superbracket on $\g$ by
\begin{equation}\label{2.1}
[X_\mu,X_\nu]=\sum_{\al=1}^n C_{\mu\nu\al} X_\al, \quad [\theta_a,X_\nu]=\sum_{b=1}^m K_{a\nu b} \theta_b, \quad [\theta_a,\theta_b]=0
\end{equation}
for some $K_{a\nu b}\in \K$. Recall that the bracket $[\, \cdot\, ,\,\cdot \,]$ is graded skew--symmetric, $[X,Y]=-(-1)^{\bar X \bar Y} [Y,X]$, and it satisfies the graded Jacobi identity
\begin{equation}\label{Jacobi}
(-1)^{\bar X \bar Z} [X,[Y,Z]]+(-1)^{\bar Y \bar X} [Y,[Z,X]]+(-1)^{\bar Z \bar Y} [Z,[X,Y]]=0
\end{equation}
for all $X,Y,Z\in \g$. This implies that the structure constants satisfy $C_{\mu\nu\al}=-C_{\nu\mu\al}$ and
\begin{align}
&\sum_{\rho=1}^n \big(C_{\mu\al\rho} C_{\rho\beta\nu}+C_{\al\beta\rho} C_{\rho\mu\nu}+C_{\beta\mu\rho} C_{\rho\al\nu}\big) = 0, \label{2.2}  \\
&\sum_{b=1}^m \big(K_{a \nu b} K_{b\mu c}-K_{a \mu b} K_{b \nu c}\big) + \sum_{\rho=1}^n C_{\mu\nu\rho} K_{a \rho c} = 0.  \label{2.3}
\end{align}
The enveloping algebra $U(\g)$ is generated by $X_\mu$ and $\theta_a$ subject to relations
\begin{align}
X_\mu X_\nu - X_\nu X_\mu &= \sum_{\al=1}^n C_{\mu\nu\al} X_\al,  \label{2.4}\\
\theta_a X_\mu - X_\mu \theta_a &= \sum_{b=1}^m K_{a\mu b} \theta_b,  \label{2.5}\\
\theta_a \theta_b + \theta_b \theta_a &= 0.  \label{2.6}
\end{align}
To simplify the notation we shall often identify the elements of $\g$ with their canonical images in $U(\g)$.
The generators $X_1, X_2, \ldots, X_n$ are interpreted as coordinates on the NC space
defined by the Lie algebra relations \eqref{2.4} and $\theta_1, \theta_2, \dots, \theta_m$ are
interpreted as one--forms on this space. In general, $n\neq m$ since the number of one--forms may not equal the number
of coordinates (see Ref. \cite{Sitarz}).

We now explain the construction of FODC of classical dimension on the enveloping algebra $U(\g_0)$. In this case we assume $n=m$.
Let $(\Delta,\epsilon, S)$ denote the primitive Hopf structure on $U(\g_0)$. The algebra $U(\g_0)$ may be considered as a quantum space for itself
since the coproduct $\Delta\colon U(\g_0)\to U(\g_0)\otimes U(\g_0)$ is both a left and right corepresentation as well as
an algebra homomorphism (for details see Ref. \cite{KS}).  Recall that a first order differential calculus over $U(\g_0)$ is a
$U(\g_0)$--bimodule $\Gamma$ together with a linear map $d\colon U(\g_0)\to \Gamma$ such that $d$ satisfies the Leibniz rule $d(XY)=(dX) Y+ X dY$ and
$\Gamma=\text{span}_{\K}\{X dY \mid X,Y\in U(\g_0)\}$. The differential calculus is said to be bicovariant if
there exists a left coaction $\Delta_L \colon \Gamma\to U(\g_0)\otimes \Gamma$ such that
\begin{equation}\label{2.1-A}
\Delta_L(X\omega Y) =\Delta(X) \Delta_L(\omega) \Delta (Y), \quad \Delta_L(X dY) = \Delta(X) (id\otimes d) \Delta(Y)
\end{equation}
and a right coaction $\Delta_R\colon \Gamma \to \Gamma\otimes U(\g_0)$ such that
\begin{equation}
\Delta_R (X \omega Y) = \Delta (X) \Delta_R(\omega) \Delta (Y), \quad \Delta_R (X dY) = \Delta(X) (d\otimes id) \Delta(Y)
\end{equation}
for all $X,Y\in U(\g_0)$, $\omega\in \Gamma$. The left and right coactions are required to satisfy the commutativity condition
\begin{equation}\label{2.5-A}
(id\otimes \Delta_R) \circ \Delta_L = (\Delta_L \otimes id) \circ \Delta_R.
\end{equation}
We define the bimodule $\Gamma=\oplus_{\al=1}^n U(\g_0)\theta_\al$
as a free left $U(\g_0)$--module with basis $\theta_1, \theta_2,\ldots, \theta_n$ and introduce the right module structure by Eq.
\eqref{2.5}. Define the linear map $d\colon U(\g_0)\to \Gamma$ by $dX_\mu = \theta_\mu$ and
extended it to $U(\g_0)$ by the Leibniz rule $d(XY)=(dX)Y+X dY$ for all $X,Y\in U(\g_0)$. Compatibility of the Leibniz rule with relation \eqref{2.4}
requires that the structure constants of $\g$ satisfy the additional condition
\begin{equation}\label{extra-condition}
K_{\mu\nu\al} - K_{\nu\mu\al} = C_{\mu\nu\al}.
\end{equation}
Define the left coaction $\Delta_L \colon \Gamma \to U(\g_0)\otimes \Gamma$ by $\Delta_L(\theta_\mu)=1\otimes \theta_\mu$ and $\Delta_L(X)=\Delta(X)$
for all $X\in U(\g_0)$ where $\Delta(X_\mu)=1\otimes X_\mu+X_\mu\otimes 1$ is the primitive coproduct.
The right coaction is defined analogously except that $\Delta_R(\theta_\mu)=\theta_\mu \otimes 1$. It is easily verified that $\Delta_L$ and $\Delta_R$ are well--defined
on $\Gamma$ since the defining relations \eqref{2.5} are in the kernels of $\Delta_L$ and $\Delta_R$. Furthermore, one easily checks that they satisfy the conditions
\eqref{2.1-A}--\eqref{2.5-A} where \eqref{2.5-A} follows from the coassociativity of the coproduct $\Delta$.

As an example, consider the $\kappa$--deformed space defined by
\begin{equation}\label{kappa}
[X_\mu,X_\nu] = i(a_\mu X_\nu - a_\nu X_\mu), \quad a_\mu \in \R, \; 1\leq \mu,\nu\leq n.
\end{equation}
This is a Lie algebra with structure constants $C_{\mu\nu\lambda}=i(a_\mu \delta_{\nu\lambda}-a_\nu \delta_{\mu\lambda})$.
It has applications in deformed (doubly)
special relativity theory \cite{KG}, quantum gravity \cite{AC} and quantum field theory \cite{G}. Differential calculi on the
$\kappa$--deformed space were classified in Ref. \cite{Meljanac-1}. There it was shown that for a given deformation vector $a\in \R^n$, $a\neq 0$,
there are three nonequivalent families of differential calculi specified by the structure constants
\begin{align}
K_{\mu\nu\lambda} &= i\frac{c}{|a|^2} a_\mu a_\nu a_\lambda - i\delta_{\mu\lambda} a_\nu, \label{S1} \\
K_{\mu\nu\lambda} &= i\frac{c}{|a|^2} a_\mu a_\nu a_\lambda - ic\delta_{\mu\lambda} a_\nu+i(1-c) \delta_{\nu\lambda} a_\mu, \\
K_{\mu\nu\lambda} &= i\frac{c}{|a|^2} a_\mu a_\nu a_\lambda - i(1+c) \delta_{\mu\nu} a_\lambda - i\delta_{\mu\lambda} a_\nu,
\end{align}
where $c\in \R$ is a free parameter. When the space deformation vanishes, i.e. $a=0$, the structure constants satisfy $\lim_{a\to 0} K_{\mu\nu\lambda}=0$.
This example demonstrates that for a given noncommutative space the differential calculus is not unique even in the classical case $n=m$.

\section{Extensions of the Lie superalgebra $\g=\g_0\oplus \g_1$ and left--right duality}
\label{section-3}

In Woronowicz's noncommutative differential calculus the commutation relations between functions and one--forms are given by an action of the Hopf
dual of the underlying Hopf algebra \cite{Woronowicz, Vladimirov}.
Our approach uses an extension of the Lie superalgebra \eqref{2.1} by a set of generators whose action on $U(\g)$ determines the commutation
relations between one--forms and monomials in $U(\g_0)$. In the present considerations we assume that $n\neq m$ in general.
At this point it is useful to introduce the following notation.
The uppercase latin letters $A,B,C$, etc. range over the indices of the entire basis of $\g$. With this convention, $\sum_A = \sum_\alpha+\sum_a$
where $\sum_{\al}$ and $\sum_a$ are the summations over the indices of $X_1, \ldots, X_n$ and $\theta_1, \ldots, \theta_m$, respectively.
Let us denote the basis of $\g$ by $Z_A$: $Z_\al=X_\al$ and $Z_{n+a}=\theta_a$. Then the commutation relations \eqref{2.4}--\eqref{2.6} can be written as
$[Z_A,Z_B]=\sum_J \mathcal{C}_{ABJ}\, Z_J$
where $[Z_A,Z_B]=Z_A Z_B-(-1)^{\bar Z_A \bar Z_B} Z_B Z_A$ and the structure constants $\mathcal C_{ABJ}$ are given by the following table:
\begin{alignat}{2}
&\mathcal C_{\mu\nu\lambda} = C_{\mu\nu\lambda}, \qquad &  &\mathcal C_{\mu\nu a} =0, \label{C-18} \\
&\mathcal C_{\mu a\nu} =0, \qquad &  &\mathcal C_{\mu ab} =-K_{a\mu b}, \\
&\mathcal C_{a \mu\nu} =0, \qquad & &\mathcal C_{a\mu b} = K_{a\mu b}, \\
&\mathcal C_{a b\mu} = 0, \qquad & &\mathcal C_{abc} = 0. \label{C-21}
\end{alignat}
Now, let $\g^L$ be the vector space over $\K$ with basis $\{Z_A, T_{AB}\mid 1\leq A,B\leq n+m\}$ where $T_{AB}$ are the elements of the block
matrix
\begin{equation}
T=\begin{bmatrix} T^{1}_{11} & \ldots & T^{1}_{1n} & T^{2}_{11} & \ldots & T^{2}_{1m} \\
\vdots & & \vdots & \vdots & & \vdots \\
T^{1}_{n1} & \ldots & T^{1}_{nn} & T^{2}_{n1} & \ldots & T^{2}_{nm} \\
0 & \ldots & 0 & T^{4}_{11} & \ldots & T^{4}_{1m} \\
\vdots & & \vdots & \vdots & & \vdots \\
0 & \ldots & 0 & T^{4}_{m1} & \ldots & T^{4}_{mm}
\end{bmatrix}
\end{equation}
(where $T^3_{a\mu}=0$). Define the degrees of $T_{AB}$ by $\overline{T_{\mu\nu}^1}=\overline{T_{ab}^4}=0$ and $\overline{T_{\mu a}^2}=1$, and
introduce the Lie superbracket on $\g^L$ by
\begin{equation}\label{2.15}
[Z_A,Z_B]=\sum_J \mathcal C_{ABJ}\, Z_J, \quad [T_{AB},T_{CD}]=0, \quad [T_{AB},Z_C]=\sum_J \mathcal C_{ACJ} T_{JB}.
\end{equation}
Explicitly, we have
\begin{alignat}{2}
[T_{\mu\nu}^{1},X_\lambda] &= \sum_{\rho} C_{\mu\lambda\rho} T_{\rho\nu}^{1}, & \quad [T_{\mu\nu}^{1},\theta_a] &=0,  \label{2.25-A}\\
[T_{\mu a}^{2},X_\lambda] &=\sum_{\rho} C_{\mu\lambda\rho} T_{\rho a}^{2}, & \quad [T_{\mu a}^{2},\theta_b] &=-\sum_{c} K_{b \mu c} T_{ca}^{4}, \label{2.26-A} \\
[T_{ab}^{4},X_\lambda] &=\sum_{c} K_{a\lambda c} T_{cb}^{4}, & \quad [T_{ab}^{4},\theta_c] &= 0.  \label{2.27-A}
\end{alignat}
Using the Jacobi identities \eqref{2.2}--\eqref{2.3} it is straightforward to verify that $\g^L=\g_0^L\oplus \g^L_1$ is a Lie superalgebra
where $\g^L_0=\text{span}_{\mathbb{K}}\{X_\mu, T_{\mu\nu}^1, T_{ab}^4\mid 1\leq \mu,\nu \leq n, 1\leq a,b\leq m\}$ and
$\g^L_1=\text{span}_{\mathbb{K}}\{\theta_a, T_{\mu a}^2 \mid 1\leq \mu \leq n, 1\leq a\leq m\}$ are the even and odd parts of $\g^L$, respectively.
The following theorem is a generalization of the result on extensions of Lie algebras given in Ref. \cite{SKJ}.

\begin{theorem}\label{tm-1}
There exists a left action $\brhd \colon U(\g^L)\otimes U(\g)\to U(\g)$, defined by
\begin{equation}
1\brhd X = X, \quad Z_A\brhd X = Z_A X, \quad T_{AB}\brhd 1 = \delta_{AB}, \quad X\in U(\g),
\end{equation}
such that
\begin{equation}\label{2.20}
T_{AB} \brhd (XY) = \sum_C (T_{AC} \brhd X) (-1)^{\overline{T}_{CB}\, \overline X} (T_{CB}\brhd Y)
\end{equation}
for all $X,Y\in U(\g)$.
\end{theorem}

\noindent\textit{Sketch of proof.} We note that the action of $T_{AB}$ on the generators $Z_C$ is determined by the commutation relations \eqref{2.15} between $T_{AB}$ and $Z_C$. Indeed, acting by
the third relation in \eqref{2.15} on $1\in U(\g)$ and using the normalization condition $T_{AB}\brhd 1=\delta_{AB}$ we find
\begin{equation}\label{2.21}
T_{AB}\brhd Z_C = (-1)^{\overline{T}_{AB}\, \overline{Z}_C} \delta_{AB} Z_C + \mathcal C_{ACB}.
\end{equation}
Explicitly,
\begin{alignat}{2}
T_{\mu\nu}^1 \brhd X_\lambda &= \delta_{\mu\nu} X_\lambda + C_{\mu\lambda\nu}, \qquad & T_{\mu\nu}^1 \brhd \theta_a &= \delta_{\mu\nu}\theta_a,  \label{2.22} \\
T_{\mu a}^2 \brhd X_\lambda &= 0, \qquad & T_{\mu a}^2 \brhd \theta_b &= -K_{b\mu a},  \label{2.23} \\
T_{ab}^4 \brhd X_\lambda &= \delta_{ab} X_\lambda + K_{a\lambda b}, \qquad & T_{ab}^4 \brhd \theta_c &= \delta_{ab} \theta_c. \label{2.24}
\end{alignat}
The action of $T_{AB}$ on higher order monomials is given by Eq. \eqref{2.20} which is proved by induction on $\deg(X)$. To this end, we need the following auxiliary
result. Applying the Leibniz rule to the supercommutator $[T_{AB},XY]$ we find
\begin{align}
[T_{AB}, XY]\brhd 1 &= \Big([T_{AB},X]\, Y +(-1)^{\overline{T}_{AB}\, \overline{X}}\, X\, [T_{AB},Y]\Big)\brhd 1 \notag \\
&=[T_{AB},X]\brhd Y + (-1)^{\overline{T}_{AB}\, \overline{X}}\, X\, (T_{AB}\brhd Y) - (-1)^{\overline{T}_{AB} \overline{XY}}\, (T_{AB}\brhd 1).  \label{2.26}
\end{align}
On the other hand, expanding $[T_{AB},XY]$ we obtain
\begin{equation}\label{2.25}
[T_{AB}, XY]\brhd 1 = T_{AB} \brhd (XY) - (-1)^{\overline{T}_{AB}\, \overline{XY}}\, XY\,(T_{AB}\brhd 1).
\end{equation}
Comparing expressions \eqref{2.26} and \eqref{2.25} we find that
\begin{equation}\label{2.27}
T_{AB}\brhd (XY) = [T_{AB},X]\brhd Y + (-1)^{\overline{T}_{AB}\, \overline{X}}\, X\, (T_{AB}\brhd Y).
\end{equation}
Now, using Eqs. \eqref{2.22}--\eqref{2.24} and Eq. \eqref{2.27} one shows by induction on $\deg(X)$ that
the action of the block elements of $[T_{AB}]$ is given by
\begin{align}
T_{\mu\nu}^1 \brhd (XY) &= \sum_{\al=1}^n (T_{\mu\al}^1 \brhd X) (T_{\al \nu}^1 \brhd Y),  \label{2.28} \\
T_{\mu b}^2 \brhd (XY) &= \sum_{a=1}^m (T_{\mu a}^2 \brhd X) (T_{ab }^4 \brhd Y) + (-1)^{\bar X}\, \sum_{\al=1}^n (T_{\mu \al}^1 \brhd X) (T_{\al b}^2 \brhd Y), \label{2.29} \\
T_{ac}^4 \brhd (XY) &= \sum_{b=1}^m (T_{ab}^4 \brhd X) (T_{bc}^4 \brhd Y).  \label{2.30}
\end{align}
It is easily verified that Eqs. \eqref{2.28}--\eqref{2.30} can be collected into the single expression \eqref{2.20}.

In the next step we prove that the
defining relations \eqref{2.4}--\eqref{2.6} of $U(\g)$ are contained in the kernel of the action $\brhd$. This is obvious for the action of $Z_A$, hence we only
consider the action of $T_{AB}$. To illustrate the point, we prove the claim for $T^2_{\lambda a}$. Using Eqs. \eqref{2.23} and \eqref{2.29} one
immediately finds
\begin{equation}
T_{\lambda a}^2 \brhd \big([X_\mu,X_\nu]-\sum_{\rho=1}^n C_{\mu \nu \rho} X_\rho\big)=0.
\end{equation}
For the defining relation \eqref{2.5} we compute
\begin{equation}
T_{\lambda b}^2 \brhd (\theta_a X_\mu) = -K_{a\lambda b} X_\mu-\sum_{c=1}^m K_{a\lambda c}\, K_{c\mu b}.
\end{equation}
Similarly,
\begin{equation}
T^2_{\lambda b} \brhd (X_\mu \theta_a) = -K_{a\lambda b} X_\mu -\sum_{\rho=1}^n C_{\lambda\mu\rho} K_{a\rho b}
\end{equation}
which yields
\begin{equation}\label{2.40}
T_{\lambda b}^2 \brhd [\theta_a,X_\mu] = \sum_{\rho=1}^n C_{\lambda\mu\rho} K_{a\rho b} -\sum_{c=1}^m K_{a\lambda c} K_{c\mu b}.
\end{equation}
Now, from Eqs. \eqref{2.40} and \eqref{2.23} we have
\begin{align}
&T_{\lambda b}^2 \brhd \Big([\theta_a,X_\mu]-\sum_{c=1}^m K_{a\mu c} \theta_c\Big) \notag \\
&=\sum_{\rho=1}^n C_{\lambda \mu \rho} K_{a\rho b}+\sum_c \big(K_{a\mu c} K_{c\lambda b}-K_{a\lambda c} K_{c\mu b}\big) = 0
\end{align}
where we used the Jacobi identity \eqref{2.3}. Finally, regarding the last defining relation \eqref{2.6} we note that
\begin{equation}
T^2_{\lambda c} \brhd (\theta_a \theta_b) = K_{b\lambda c} \theta_a - K_{a\lambda c} \theta_b
\end{equation}
which implies $T^2_{\lambda c} \brhd [\theta_a,\theta_b]=0$. Similar arguments apply to the actions of $T^1_{\mu\nu}$ and $T^4_{ab}$.
In the final step it is necessary to prove that the action $\brhd$ is also consistent with relations
\eqref{2.15}, i.e. that
\begin{equation}
[T_{AB},T_{CD}]\brhd X=0 \quad \text{and}\quad \big([T_{AB},Z_C]-\sum_J \mathcal C_{ACJ} T_{JB}\big)\brhd X = 0
\end{equation}
for all $X\in U(\g)$. This is done by induction on $\deg(X)$ using relations \eqref{2.28}--\eqref{2.30}. \qed\\

The action of $T_{AB}$ defined in Theorem \ref{tm-1} gives the commutation relations between the generators of $U(\g)$ and elements of the subalgebra $U(\g_0)$.
This is a consequence of the following result.

\begin{proposition}\label{prop-01}
For any $X\in U(\g_0)$ we have
\begin{equation}\label{2.44}
Z_A X = \sum_B (T_{AB}\brhd X) Z_B.
\end{equation}
\end{proposition}

\proof We prove the claim by induction on $\deg(X)$. Let $X=X_\mu$. If $Z_A=X_\al$, then using the first two relations in \eqref{2.22} and \eqref{2.23}
one finds that $X_\al X_\mu = \sum_B (T_{\al B}\brhd X_\mu) Z_B$. Similarly,
if $Z_A=\theta_a$, the first relation in \eqref{2.24} yields $\theta_a X_\mu = \sum_B (T_{aB}\brhd X_\mu) Z_B$ where we have taken into account that $T_{a\beta}^3=0$.
Assume that Eq. \eqref{2.44} holds for all monomials $X$ with $\deg(X)\leq n$. Then, in view of Theorem \ref{tm-1} we have
\begin{align}
Z_A (X_\mu X) &= \Big(\sum_B (T_{AB}\brhd X_\mu) Z_B\Big) X = \sum_B (T_{AB}\brhd X_\mu) \sum_C (T_{BC}\brhd X) Z_C  \notag \\
&= \sum_C \Big[\sum_B (T_{AB}\brhd X_\mu) (T_{BC}\brhd X)\Big] Z_C = \sum_C \Big(T_{AC}\brhd (X_\mu X)\Big)Z_C.
\end{align}
Hence, Eq. \eqref{2.44} holds for all monomials $X\in U(\g_0)$ and extends linearly to $U(\g_0)$. \qed\\

In the special case when $Z_A=\theta_a$ Eq. \eqref{2.44} yields
\begin{equation}\label{2.46}
\theta_a X = \sum_{b=1}^m (T_{ab}^4 \brhd X) \theta_b, \quad X\in U(\g_0).
\end{equation}
We note that $T^4_{ab}\brhd X=\delta_{ab}X+p_{ab}(X)$ where $p_{ab}(X)\in U(\g_0)$ is a polynomial with $\deg(p_{ab}(X))<\deg (X)$.
Thus, the action of $T^4_{ab}$ gives the commutation relations between one--forms and functions on the quantum space $U(\g_0)$.

Relation \eqref{2.44} can also be interpreted as follows. If $X$ is a monomial in $U(\g_0)$, then shifting $Z_A$ to the far right in the product $Z_A X$
generates polynomials $Q_{AB}(X)\in U(\g_0)$ such that $Z_A X = \sum_B Q_{AB}\, Z_B$ where $Q_{AB}(X)=T_{AB}\brhd X$. Now consider the dual
problem. If $Z_A$ is shifted to the far left in $X Z_A$, this generates polynomials $\widetilde Q_{BA}(X)$ such that $X Z_A=\sum_B Z_B\, \widetilde Q_{BA}(X)$.
One can show that the Lie superalgebra $\g$ can be extended by operators $S_{AB}$ such that $\widetilde Q_{BA}(X)=S_{AB}\brhd X$.
This extension, denoted $\g^R$, is defined by
\begin{equation}
[S_{AB},S_{CD}]=0, \quad [S_{AB}, Z_C] = -\sum_J \mathcal C_{JCB}\, S_{AJ}.
\end{equation}
In analogy with Theorem \ref{tm-1}, the enveloping algebra $U(\g^R)$ acts on $U(\g)$ by $1\brhd X=X$, $Z_A\brhd X = Z_A X$, $S_{AB}\brhd 1 =
\delta_{AB}$ and
\begin{equation}
S_{AB}\brhd (XY) = \sum_C (S_{CB}\brhd X) (-1)^{\bar S_{AC}\bar X} (S_{AC}\brhd Y)
\end{equation}
for all $X,Y\in U(\g)$. Furthermore, for any $X\in U(\g_0)$, the shift of $Z_A$ to the far left is given by
\begin{equation}
X Z_A = \sum_B Z_B (S_{AB}\brhd X).
\end{equation}
In view of this property, we refer to $T_{AB}$ and $S_{AB}$ as the left and right shift operators, respectively. We remark that the algebra $U(\g)$ can be extended by both sets
of generators $T_{AB}$ and $S_{AB}$ such that
\begin{equation}
\sum_J S_{AJ} T_{JB} = \sum_J T_{AJ} S_{JB}=\delta_{JB}.
\end{equation}
Hence, $S=[S_{AB}]$ is the formal inverse of the matrix $T=[T_{AB}]$.

We remark that the operators $T_{AB}$ and $S_{AB}$ do not have a  direct physical meaning. Rather,
they provide a tool for computing the commutation relations between forms and functions on the underlying NC space. They are also used to write a given monomial
$X\in U(\g_0)$ in the PBW basis using Eq. \eqref{2.44} which facilitates computation in the extended algebra $U(\g)$.
A realization of $T_{AB}$ in terms of differential operators is presented in Section \ref{section-4}.

\section{Realizations of the Lie superalgebra $\g=\g_0\oplus \g_1$}
\label{section-4}

In this section we consider realizations of the Lie superalgebra $\g=\g_0\oplus\g_1$ by formal power series in a semicompleted Weyl superalgebra
(Clifford--Weyl algebra). These realizations are used in Section \ref{section-5} to obtain the bicovariant calculus from Section
\ref{section-2} as a deformation of the standard differential calculus on the Euclidean space $\R^n$.
The Weyl superalgebra $\A_{(n,m)}$ is generated by four sets of generators $x_\mu,\p_\mu, \xi_a, q_a$, $1\leq \mu\leq n$, $1\leq a\leq m$,
with degrees  defined by $\bar x_\mu = \bar \p_\mu=0$ and $\bar \xi_a = \bar q_a =1$. The supercommutator on $\A_{(n,m)}$ is given by $[x,y]=xy-(-1)^{\bar x \bar y} yx$
and the generators are subject to relations
\begin{equation}\label{sc}
[\p_\mu,x_\nu]=\delta_{\mu\nu}, \quad [q_a,\xi_b]=\delta_{ab}
\end{equation}
with all other relations between the generators being zero. The supercommutator satisfies the graded Jacobi identity \eqref{Jacobi}.
Let $\hat \A_{(n,m)}$ denote the semicompletion of $\A_{(n,m)}$ which contains formal power series in $\p_\mu$ and $q_a$,
but only polynomial expressions in $x_\mu$ and $\xi_a$. First, we recall some
relevant result from Refs. \cite{SKJ} and \cite{Durov}. According to Eq. \eqref{2.1} the Lie algebra $\g_0$ is defined by the commutation relations
$[X_\mu,X_\nu]=\sum_{\al=1}^n C_{\mu\nu\al} X_\al$.
Let $\CC$ be the matrix given by $\CC_{\mu\nu}=\sum_{\al=1}^n C_{\mu\al\nu}\, \p_\al$ and let $\psi(t)$ be the generating function for the Bernoulli numbers $B_k$,
\begin{equation}
\psi(t)=\frac{t}{1-e^{-t}}=\sum_{k=0}^\infty B_k \frac{(-1)^k}{k!} t^k.
\end{equation}
Then the map
\begin{equation}\label{3.55}
X_\mu \mapsto \hat x_\mu=\sum_{\al=1}^n x_\al \psi(\CC)_{\mu\al}
\end{equation}
is a realization of $\g_0$. It is called the Weyl symmetric realization since it
corresponds to the Weyl symmetric ordering on the enveloping algebra $U(\g_0)$. For the $\kappa$--deformed space defined by Eq. \eqref{kappa}
examples of different realizations and their applications in physics can be found in Refs. \cite{Meljanac-5,Meljanac-6}, and for more general
noncommutative spaces in Ref. \cite{Meljanac-7}.

The following is a generalization of the above result to the Lie superalgebra
$\g=\g_0\oplus \g_1$ defined by Eq. \eqref{2.1}.

\begin{proposition}\label{prop-02}
Define a linear map $\varphi \colon \g\to \hat \A_{(n,m)}$ on the basis of $\g$ by
\begin{equation}\label{3.56}
\varphi(X_\mu) = \sum_{\al=1}^n x_\al \psi(\CC)_{\mu\al}-\sum_{a,b=1}^m K_{b\mu a} \xi_a q_b, \quad \varphi(\theta_a)=\xi_a.
\end{equation}
Then $\varphi$ is a realization of $\g$.
\end{proposition}

\proof Let us denote $\hat x_\mu = \varphi(X_\mu)$ and $\hat \theta_a = \varphi (\theta_a)$. Obviously, $[\hat \theta_a, \hat \theta_b]=[\xi_a, \xi_b]=0$.
According to Eq. \eqref{3.55}, the Weyl symmetric realization of $\g_0$ is given by  $\hat x_\mu^\prime = \sum_{\al=1}^n x_\al \psi(\CC)_{\mu\al}$, hence
$[\hat x_\mu^\prime,\hat x_\nu^\prime]=\sum_{\rho=1}^n C_{\mu\nu\rho}\, \hat x_\rho^\prime$. This implies
\begin{align}\label{3.57}
[\hat x_\mu, \hat x_\nu] &= \Big[\hat x_\mu^\prime-\sum_{a,b=1}^m K_{b\mu a}\, \xi_a q_b, \hat x_\nu^\prime - \sum_{k,l=1}^m K_{l\nu k}\, \xi_k q_l\Big] \\
&=\sum_{\rho=1}^n C_{\mu\nu\rho}\, \hat x_\rho^\prime + \sum_{a,b=1}^m \Big(\sum_{l=1}^m (K_{b\nu l} K_{l\mu a}-K_{b\mu l} K_{l\nu a})\Big) \xi_a q_b.
\end{align}
Using the Jacobi identity \eqref{2.3} we find
\begin{equation}
[\hat x_\mu, \hat x_\nu] = \sum_{\rho=1}^n C_{\mu\nu\rho} \Big(\hat x_\rho^\prime - \sum_{a,b=1}^m K_{b\rho a}\, \xi_a q_b\Big)=\sum_{\rho=1}^n C_{\mu\nu\rho}\, \hat x_\rho.
\end{equation}
Furthermore, one easily verifies that
\begin{equation}
[\hat \theta_a, \hat x_\mu] = \Big[\xi_a, \hat x_\mu^\prime-\sum_{k,l=1}^m K_{l\mu k}\, \xi_k q_l\Big] = \sum_{k=1}^m K_{a\mu k}\, \hat \theta_k.
\end{equation}
Thus, $\hat x_\mu$ and $\hat \theta_a$ satisfy relations \eqref{2.1}.
\qed\\

We refer to Eq. \eqref{3.56} as the Weyl--linear realization of $\g$. This realization can be extended to the Lie superalgebra $\g^L$ which includes
the shift operators $T_{AB}$ defined by Eq. \eqref{2.15}.

\begin{theorem}\label{tm-2}
Let $\varphi \colon \g^L\to \hat \A_{(n,m)}$ be a linear map defined on the basis of $\g^L$ by
\begin{align}
&\varphi(X_\mu) = \sum_{\al=1}^n x_\al \psi(\CC)_{\mu\al}-\sum_{a,b=1}^m K_{b\mu a} \xi_a q_b, \quad \varphi (\theta_a)=\xi_a, \label{3.61} \\
&\varphi(T_{\mu\nu}^1) = (e^\CC)_{\mu\nu}, \quad \varphi(T_{\mu a}^2)=-\sum_{b,c=1}^m K_{b\mu c}\, (e^\KK)_{ca}\, q_b, \quad \varphi(T_{ab}^4)=(e^\KK)_{ab} \label{3.62}
\end{align}
where $\KK$ is the matrix $\KK_{ab}=\sum_{\rho=1}^n K_{a\rho b} \p_\rho$. Then $\varphi$ is a realization of $\g^L$.
\end{theorem}

\proof Denote $\hat T^1_{\mu\nu}=\varphi(T^1_{\mu\nu})$, $\hat T^2_{\mu a} = \varphi(T^2_{\mu a})$ and $\hat T^4_{ab}=\varphi(T^4_{ab})$. In view of
Proposition \ref{prop-02}, the restriction $\varphi\mid_{\g}\colon \g\to \hat \A_{(n,m)}$ is a realization of the Lie superalgebra $\g$. Since the realizations
of the shift operators clearly commute, $[\hat T_{AB},\hat T_{CD}]=0$, it remains to prove that $\hat T_{AB}$ and $\hat Z_C$ satisfy relations
\eqref{2.25-A}--\eqref{2.27-A}. Denote $\hat x_\lambda^\prime = \sum_{\al=1}^n x_\al \psi(\CC)_{\lambda\al}$. It was shown in Ref. \cite{SKJ}
(cf. Theorem 3) that $\hat x^\prime_{\lambda}$ and $\hat T^1_{\mu\nu}$
satisfy $[\hat T^1_{\mu\nu},\hat x_\lambda^\prime]=\sum_{\rho=1}^n C_{\mu\lambda\rho}\, \hat T^1_{\rho\nu}$. This implies
\begin{equation}
[\hat T^1_{\mu\nu}, \hat x_\lambda]=\Big[\hat T^1_{\mu\nu}, \hat x^\prime_\lambda -\sum_{a,b=1}^m K_{b\mu a}\, \xi_a q_b\Big]=\sum_{\rho=1}^n C_{\mu\lambda\rho}\,
\hat T^1_{\rho\nu}.
\end{equation}
Furthermore,
\begin{equation}
[\hat T^1_{\mu\nu}, \hat \theta_a] = \big[(e^\CC)_{\mu\nu}, \xi_a\big] = 0
\end{equation}
since $[\p_\mu,\xi_a]=0$. Regarding the shift operator $\hat T^4_{ab}$, we have
\begin{equation}
[\hat T^4_{ab}, \hat x_\lambda] = \sum_{\rho=1}^n \big[(e^\KK)_{ab}, x_\rho \psi(\CC)_{\lambda\rho}\big]=\sum_{\rho=1}^n \frac{\p}{\p \p_\rho} (e^\KK)_{ab}\, \psi(\CC)_{\lambda\rho}.
\end{equation}
It is proved in Lemma \ref{lm-1} (see Appendix) that the above expression reduces to
\begin{equation}
\sum_{\rho=1}^n \frac{\p}{\p \p_\rho} (e^\KK)_{ab}\, \psi(\CC)_{\lambda\rho} = \sum_{c=1}^m K_{a\lambda c}\, (e^\KK)_{cb}.
\end{equation}
As a result, we find that
\begin{equation}\label{3.67}
[\hat T^4_{ab}, \hat x_\lambda] = \sum_{c=1}^m K_{a\lambda c}\, \hat T^4_{cb}.
\end{equation}
In addition,
\begin{equation}
[\hat T^4_{ab}, \hat \theta_c] = \big[(e^\KK)_{ab}, \xi_c\big] = 0.
\end{equation}
To find the commutation relations for $\hat T^2_{\mu a}$, we compute
\begin{equation}\label{3.69}
[\hat T_{\mu a}^2, \hat x_\lambda] = -\sum_{c,d=1}^m K_{d\mu c} \Big((e^\KK)_{ca}\, [q_d,\hat x_\lambda]+\big[(e^\KK)_{ca}, \hat x_\lambda\big] q_d\Big).
\end{equation}
Using the relation $[q_d, \hat x_\lambda]=-\sum_{b=1}^m K_{b\lambda d}\, q_b$ and substituting Eq. \eqref{3.67} into Eq. \eqref{3.69} we obtain
\begin{equation}\label{3.70}
[\hat T^2_{\mu a}, \hat x_\lambda] = \sum_{b,c=1}^m \sum_{d=1}^m (K_{b\lambda d} K_{d\mu c} - K_{b\mu d} K_{d\lambda c}) (e^\KK)_{ca}\, q_b.
\end{equation}
It follows from the Jacobi identity \eqref{2.3} that the above expression can be written as
\begin{equation}
[\hat T^2_{\mu a}, \hat x_\lambda] = \sum_{\rho=1}^n C_{\mu\lambda\rho}\, \Big(-\sum_{b,c=1}^m K_{b\rho c}\, (e^\KK)_{ca}\, q_b\Big)=\sum_{\rho=1}^n C_{\mu\lambda\rho}\,
\hat T^2_{\rho a}.
\end{equation}
Finally, the remaining relation yields
\begin{equation}
[\hat T^2_{\mu a}, \hat \theta_a] = -\sum_{c,d=1}^m K_{d\mu c}\, (e^\KK)_{ca}\, [q_d,\xi_b]=-\sum_{c=1}^m K_{b\mu c}\, \hat T^4_{ca}.
\end{equation}
This completes the proof that \eqref{3.61}--\eqref{3.62} is a realization of $\g^L$. \qed\\

Let us consider a simple example of differential calculus of classical dimension $(n=m)$ on the $\kappa$--deformed space defined by Eq. \eqref{kappa}.
(Recall that when $n=m$, the indices of $\theta_1, \theta_2,\ldots, \theta_n$ are denoted by lowercase greek letters). In the special case $c=0$ in Eq. \eqref{S1} we have
$K_{\mu\nu\lambda}=-i\delta_{\mu\lambda} a_\nu$. This corresponds to the algebra S1 in Ref. \cite{Meljanac-1},
\begin{equation}\label{S1-kappa}
[\theta_\mu,X_\nu]=-ia_\nu \theta_\mu.
\end{equation}
It was shown in Ref. \cite{SKJ} (see also Ref. \cite{Meljanac-6}) that the Weyl symmetric realization of the Lie algebra \eqref{kappa}
is given by
\begin{equation}
\hat x_\mu^\prime=\sum_{\al=1}^n x_\al \psi(\CC)_{\mu\al} = x_\mu \frac{A}{e^A-1}+ia_\mu (x\cdot\p) \left(\frac{1}{A}-\frac{1}{e^A-1}\right)
\end{equation}
where $A=i\sum_{\nu=1}^n a_\nu \p_\nu$. Therefore, the Weyl--linear realization \eqref{3.61} yields
\begin{equation}
\hat x_\mu = \hat x_\mu^\prime -\sum_{\al,\beta=1}^n K_{\beta\mu\al} \xi_\al q_\beta =x_\mu \frac{A}{e^A-1}+ia_\mu (x\cdot\p)
\left(\frac{1}{A}-\frac{1}{e^A-1}\right)+ia_\mu \sum_{\al=1}^n \xi_\al q_\al.
\end{equation}
It was also shown in Ref. \cite{SKJ} that the shift operator $\hat T_{\mu\nu}^1$ is given by
\begin{equation}
\hat T_{\mu\nu}^1 = \delta_{\mu\nu}\, e^{-A} -ia_\mu \p_\nu \frac{e^A-1}{A}.
\end{equation}
In order to find the realizations of $T_{\mu\nu}^2$ and $T_{\mu\nu}^4$ we note that $e^\KK=e^{-A}I$ where $I$ is the $n\times n$ unit matrix. Hence,
using Eq. \eqref{3.62} we find
\begin{equation}
\hat T_{\mu\nu}^2 = ia_\mu\, e^{-A} q_\nu, \quad \hat T_{\mu\nu}^4 = \delta_{\mu\nu}\, e^{-A}.
\end{equation}

\subsection{Similarity transformation and Weyl symmetric realization}

Different realizations of Lie superalgebras can be obtained by using automorphisms of the algebra $\hat \A_{(n,m)}$.
In this section we consider a particular example of such transformation that leads to an important conjecture about
realizations of the Lie superalgebras defined by Eq. \eqref{2.1}. Let $\tilde x_\mu, \tilde \p_\mu, \tilde \xi_a, \tilde q_a$, $1\leq
\mu\leq n$, $1\leq a\leq m$, denote a set of generators of the Weyl superalgebra $\hat \A_{(n,m)}$. Let $\tilde \CC$ and $\tilde
\KK$ be the matrices defined by $\tilde \CC_{\mu\nu}=\sum_{\rho=1}^n C_{\mu\rho\nu}\tilde \p_\rho$ and
$\tilde \KK_{ab}=\sum_{\rho=1}^n K_{a\rho b} \tilde \p_\rho$. Consider an automorphism $\phi\colon \hat \A_{(n,m)}\to \hat
\A_{(n,m)}$ defined by $\phi(u)=S u S^{-1}$ where $S=\exp\big(\sum_{a,b=1}^m \tilde \xi_a \tilde q_b\, (\ln M)_{ba}\big)$ and $M$ is the matrix given by
\begin{equation}
M=\psi(\tilde \KK) = \frac{\tilde \KK}{1-e^{-\tilde \KK}}.
\end{equation}
Let $x_\mu, \p_\mu, \xi_a, q_a$ be the images of the generators $\tilde x_\mu, \tilde \p_\mu, \tilde \xi_a, \tilde q_a$ under the
transformation $\phi$. It is shown in Lemma \ref{lm-2} (see Appendix) that the transformation is explicitly given by
\begin{alignat}{2}
x_\mu &= \tilde x_\mu+\sum_{a,b=1}^m \tilde \xi_a \tilde q_b\Big( M^{-1} \frac{\p M}{\p \tilde \p_\mu}\Big)_{ba}, & \quad \p_\mu &= \tilde \p_\mu, \label{4.87} \\
\xi_a &= \sum_{b=1}^m \tilde \xi_b M_{ab}, & \quad q_a &= \sum_{b=1}^m \tilde q_b M^{-1}_{ba}.  \label{4.88}
\end{alignat}
Since $\phi$ is an automorphism, the elements $x_\mu, \p_\mu, \xi_a, q_b$, $1\leq \mu\leq n$, $1\leq a\leq m$, are also
generators of $\hat \A_{(n,m)}$ satisfying the same defining relations. Hence, by substituting expressions
\eqref{4.87}--\eqref{4.88} into Eq. \eqref{3.56} we obtain a new realization of the Lie superalgebra $\g$:
\begin{align}
\varphi (X_\mu) &= \sum_{\al=1}^n \tilde x_\al \psi(\tilde \CC)_{\mu\al} + \sum_{a,b=1}^m \tilde \xi_a \tilde q_b H_{b\mu a}(\tilde \p),  \label{4.89} \\
\varphi (\theta_a) &= \sum_{b=1}^m \tilde \xi_b M_{ab}, \label{4.90}
\end{align}
where
\begin{equation}\label{4.91}
H_{b\mu a}(\tilde \p) = \sum_{c=1}^m M^{-1}_{bc}\Big(\sum_{\rho=1}^n \frac{\p M_{ca}}{\p \tilde \p_\rho}\, \psi(\tilde \CC)_{\mu\rho}-\sum_{d=1}^m K_{c\mu d}\, M_{da}\Big).
\end{equation}
It is interesting to note that there is a close relationship between the realization \eqref{4.89}--\eqref{4.90} and the extension
of the Weyl symmetric formula \eqref{3.55} to the Lie superalgebra $\g$ defined by Eq. \eqref{2.1}. To see this, denote $\tilde D=(\tilde
\p_1, \ldots, \tilde \p_n, \tilde q_1, \ldots, \tilde q_m)$ and $\tilde z = (\tilde x_1, \ldots, \tilde x_n, \tilde \xi_1, \ldots,
\tilde \xi_m)$, and let $\tilde{\mathcal{C}}$ be the $(n+m)\times (n+m)$ matrix defined by
\begin{equation}
\tilde{\mathcal{C}}_{AB} = \sum_J \mathcal{C}_{AJB} \tilde D_J
\end{equation}
where $\mathcal{C}_{AJB}$ are the structure constants of $\g$ given by \eqref{C-18}--\eqref{C-21}. By analogy with Eq. \eqref{3.55}, let us define a linear map
$\tilde \varphi \colon \g \to \hat \A_{(n,m)}$ on the basis of $\g$ by
\begin{equation}\label{4.97}
\tilde \varphi(Z_A) = \sum_B \tilde z_B\, \psi(\tilde{\mathcal{C}})_{AB}.
\end{equation}
It is fairly difficult to show by direct computation that $\tilde \varphi$ is a
realization of $\g$, but an argument in favor of this conjecture can be given as follows. The matrix $\tilde{\mathcal{C}}$ has the block--triangular form
\begin{equation}
\tilde{\mathcal{C}} = \begin{pmatrix} \tilde{\CC} & \tilde{\mathbf{L}} \\ \mathbf{0} & \tilde{\KK}\end{pmatrix}
\end{equation}
where $\tilde{\mathbf{L}}_{\mu a}=-\sum_{b=1}^m K_{b\mu a}\tilde q_b$. Hence, the matrix $\psi(\tilde{\mathcal{C}})$ is given by
\begin{equation}\label{4.94}
\psi(\tilde{\mathcal{C}})=\begin{pmatrix} \psi(\tilde{\CC}) & \tilde{\mathbf{F}} \\ \mathbf{0} & \psi(\tilde{\KK})\end{pmatrix}
\end{equation}
where
\begin{equation}\label{4.94-A}
\tilde{\mathbf{F}}=\sum_{k=1}^\infty \sum_{l=1}^k \frac{(-1)^k}{k!} B_k\, \tilde{\CC}^{k-l}\, \tilde{\mathbf{L}}\, \tilde{\KK}^{l-1}.
\end{equation}
The elements of the matrix $\tilde{\mathbf{F}}$ are of the form
\begin{equation}\label{4.96}
\tilde{\mathbf{F}}_{\mu a} = \sum_{b=1}^m \tilde q_b P_{b\mu a}(\tilde \p)
\end{equation}
where $P_{b\mu a}(\tilde \p)$ is a formal power series depending on the structure constants $C_{\mu\nu\lambda}$ and $K_{a\mu b}$. Substituting Eqs. \eqref{4.94}
and \eqref{4.96} into Eq. \eqref{4.97}  we find that $\tilde \varphi (Z_A)$ is explicitly given by
\begin{align}
\tilde \varphi (X_\mu) &= \sum_{\al=1}^n \tilde x_\al\, \psi(\tilde{\CC})_{\mu\al}+\sum_{a,b=1}^m \tilde \xi_a\, \tilde q_b P_{b\mu a}(\tilde \p),  \label{4.98} \\
\tilde \varphi (\theta_a) &= \sum_{b=1}^m \tilde \xi_b\,
\psi(\tilde \KK)_{ab}.  \label{4.99}
\end{align}
Comparing Eqs. \eqref{4.89}--\eqref{4.90} with \eqref{4.98}--\eqref{4.99} we see that $\varphi(\theta_a)=\tilde \varphi(\theta_a)$ since $M_{ab}=\psi(\KK)_{ab}$, and
$\varphi(X_\mu)=\tilde \varphi(X_\mu)$ provided $H_{b\mu a}(\tilde \p)=P_{b\mu a}(\tilde \p)$.  Since it is difficult to prove in full
generality that $H_{b\mu a} = P_{b\mu a}$, we consider the power series expansion for $H_{b\mu a}$ and $P_{b\mu a}$. A lengthly but
straightforward computation using Eq. \eqref{4.91} shows that the first order approximation of $H_{b\mu a}$ is given by
\begin{equation}
H_{b\mu a}(\tilde \p) \approx -\frac{1}{2}K_{b\mu a}+\sum_{\al=1}^n \left(-\frac{5}{12}\sum_{c=1}^m K_{b\mu c} K_{c\al a}+
\frac{1}{3} \sum_{c=1}^m K_{b\al c} K_{c\mu a}+\frac{1}{4}\sum_{\rho=1}^n C_{\mu\al\rho} K_{b\rho a}\right) \tilde \p_\al.
\end{equation}
Using the Jacobi identity \eqref{2.3}, the above approximation can be transformed into a simplified form
\begin{equation}\label{4.101}
H_{b\mu a}(\tilde \p) \approx -\frac{1}{2}K_{b\mu a} - \frac{1}{12}\sum_{\al=1}^n \left(\sum_{\rho=1}^n C_{\mu\al\rho}
K_{b\rho a}+\sum_{c=1}^m K_{b\mu c} K_{c\al a}\right) \tilde \p_\al.
\end{equation}
On the other hand, approximation of the matrix $\tilde{\mathbf{F}}$ in Eq. \eqref{4.94-A}  to first order in $\tilde \p_\al$ yields
\begin{equation}
\tilde{\mathbf{F}} \approx \sum_{b=1}^m q_b \left[-\frac{1}{2}K_{b\mu a} - \frac{1}{12}\sum_{\al=1}^n
\left(\sum_{\rho=1}^n C_{\mu\al\rho} K_{b\rho a}+\sum_{c=1}^m K_{b\mu c} K_{c\al a}\right) \tilde \p_\al\right],
\end{equation}
hence $P_{b\mu a}=H_{b\mu a}$ to first order in $\tilde \p_\al$. Based on this computation we conjecture that Eq. \eqref{4.97} is indeed a realization
of $\g$ obtained by extending the Weyl symmetric realization of $\g_0$ to the Lie superalgebras defined by Eq. \eqref{2.1}.
A full proof of this conjecture will be considered elsewhere.

\section{Relization of the exterior derivative in the special case $n=m$}
\label{section-5}

In this section we construct the bicovariant differential calculus from Section \ref{section-2} as a deformation of the standard differential
calculus on the Euclidean space $\R^n$. Here we assume that $n=m$, i.e. the number of coordinates $X_\mu$ is equal to the number
of one--forms $\theta_\mu$. Note that in this case the condition \eqref{extra-condition} holds. Starting with the Weyl--linear realization \eqref{3.56} we want to find
a realization of the exterior derivative $d\colon U(\g_0)\to \Gamma$. First, let us rephrase the differential calculus on $\R^n$ in a more algebraic language.
In the undeformed case, $U(\g_0)$ is simply the algebra $A\subset \hat \A_{(n,n)}$ generated by commutative coordinates $x_1,x_2, \ldots, x_n$.
Since $\xi_\mu \xi_\nu = -\xi_\nu \xi_\mu$, the generators $\xi_\mu$ are interpreted as one--forms on $\R^n$. The algebra of differential
forms is a graded algebra $\Omega=\oplus_{k=0}^n \Omega^k$ where $\Omega^0=A$
and $\Omega^k=\text{span}\big\{f\, \xi_{\mu_1} \xi_{\mu_2} \ldots \xi_{\mu_k}  \mid f\in A,\; 1\leq \mu_1<\mu_2 \ldots <\mu_k\leq n\big\}$.
The exterior derivative is a degree--one map $d\colon \Omega^k \to \Omega^{k+1}$ such that
$d^2=0$ and $d(\omega \eta) = (d\omega) \eta + (-1)^k \omega\, d\eta\,$ for $\omega\in \Omega^k$ and $\eta\in \Omega$. For zero--forms, the map $d\colon \Omega^0\to \Omega^1$
is realized as $d=[\sum_{\al=1}^n \xi_\al \p_\al, \, \cdot\, ]$ since $df=\big[\sum_{\al=1}^n \xi_\al \p_\al, f\big]=\sum_{\al=1}^n \frac{\p f}{\p x_\al}\, \xi_\al$
(where $[\, \cdot\, ,\, \cdot\, ]$ is defined by Eq. \eqref{sc}). This suggests that for a given realization of
$X_\mu$ and $\theta_\mu$ we should define $d\colon U(\g_0)\to \Gamma$ as a map $d=[\hat d, \, \cdot \,]$ for some element $\hat d\in \hat \A_{(n,n)}$.
Since $\hat d$ is a deformation of $\hat d_0 = \sum_{\al=1}^n \xi_\al \p_\al$, we assume that
\begin{equation}\label{3.70-A}
\hat d = \sum_{\al=1}^n \xi_\al \Lambda_\al(\p)
\end{equation}
where $\Lambda_\al(\p) = \p_\al + O(\p^2)$ is a formal power series in $\p_1, \p_2, \ldots, \p_n$ (if the structure constants of $\g$ vanish, then $\Lambda_\al(\p)=\p_\al$).
The function $\Lambda_\al(\p)$ is uniquely
determined by the condition $d\hat x_\mu = [\hat d, \hat x_\mu]=\hat \theta_\mu$. Using the realization \eqref{3.56} (with $n=m$) we find
\begin{equation}
[\hat d, \hat x_\mu] = \sum_{\al=1}^n \Big[\sum_{\beta=1}^n \frac{\p \Lambda_\al}{\p\, \p_\beta}\,\psi(\CC)_{\mu\beta} + \sum_{\beta=1}^n K_{\beta\mu\al} \Lambda_\beta(\p)\Big].
\end{equation}
The condition $[\hat d,\hat x_\mu]=\hat \theta_\mu$ implies that the functions $\Lambda_\al(\p)$ satisfy a system of partial differential equations
\begin{equation}\label{3.75}
\sum_{\beta=1}^n \frac{\p \Lambda_\al}{\p\, \p_\beta}\, \psi(\CC)_{\mu\beta}+\sum_{\beta=1}^n K_{\beta\mu\al} \Lambda_\beta(\p)=\delta_{\al\mu}
\end{equation}
with initial condition $\Lambda_\al(0)=0$. Multiplying Eq. \eqref{3.75} by $\p_\mu$ and summing over $\mu$ we obtain
\begin{equation}\label{3.76}
\sum_{\beta=1}^n \frac{\p \Lambda_\al}{\p\, \p_\beta} \Big(\sum_{\mu=1}^n \psi(\CC)_{\mu\beta}\, \p_\mu\Big)+\sum_{\beta=1}^n \KK_{\beta\al} \Lambda_\beta(\p) = \p_\al
\end{equation}
where $\KK_{\beta\al} = \sum_{\mu=1}^n K_{\beta\mu\al} \p_\mu$. Since the structure constants are antisymmetric, $C_{\mu\nu\al} = - C_{\nu\mu\al}$, one easily shows
by induction on $k$ that $\sum_{\mu=1}^n (\CC^k)_{\mu\beta}\, \p_\mu =0$ for all $k\geq 1$. Hence, expanding $\psi(\CC)$ into power series we find $\sum_{\mu=1}^n \psi(\CC)_{\mu\beta}\, \p_\mu
=\p_\beta$. Therefore, Eq. \eqref{3.76} takes the form
\begin{equation}\label{3.77}
\sum_{\beta=1}^n \frac{\p \Lambda_\al}{\p\, \p_\beta}\, \p_\beta + \sum_{\beta=1}^n \KK_{\beta\al}\, \Lambda_\beta(\p) = \p_\al.
\end{equation}
We note that the differential equation \eqref{3.77} is obtained from
\begin{equation}\label{3.78}
\frac{d}{dt} \Lambda_\al (t \p) + \sum_{\beta=1}^n \KK_{\beta\al}\, \Lambda_\beta(t \p)=\p_\al
\end{equation}
when the left--hand side of \eqref{3.78} is evaluated at $t=1$. Hence, it suffices to find a solution of Eq. \eqref{3.78} satisfying $\Lambda_\al(0)=0$.
We look for the solution as a power series in $t$,
\begin{equation}\label{3.79}
\Lambda_\al(t \p) = \sum_{k=0}^\infty \Lambda_\al^{(k)} (\p)\, t^k,
\end{equation}
and substitute it into Eq. \eqref{3.78} to find a recurrence relation for $\Lambda_\al^{(k)}$:
\begin{equation}
\Lambda_\al^{(k+1)} = -\frac{1}{k+1} \sum_{\beta=1}^n \KK_{\beta\al}\, \Lambda_\beta^{(k)}, \quad k\geq 1,
\end{equation}
where $\Lambda_\beta^{(0)}=0$ and $\Lambda_\beta^{(1)}=\p_\beta$. The solution of the recurrence relation
is given by
\begin{equation}
\Lambda_\al^{(k)} = \frac{(-1)^{k-1}}{k!} \sum_{\beta=1}^n (\KK^{k-1})_{\beta\al}\, \p_\beta, \quad k\geq 1.
\end{equation}
As a result, the solution of Eq. \eqref{3.78} is given by
\begin{equation}\label{3.82}
\Lambda_\al(t \p) = \sum_{\beta=1}^n \Big[\sum_{k=1}^\infty \frac{(-1)^{k-1}}{k!} (t \KK)^{k-1}\Big]_{\beta\al} (t \p_\beta)
=\sum_{\beta=1}^n \Big(\frac{1-e^{-t\KK}}{\KK}\Big)_{\beta\al} \p_\beta.
\end{equation}
Evaluating Eq. \eqref{3.82} at  $t=1$ and substituting the resulting expression into Eq. \eqref{3.70-A} we obtain
\begin{equation}
\hat d = \sum_{\al,\beta=1}^n \Big(\frac{1-e^{-\KK}}{\KK}\Big)_{\beta\al} \xi_\al \p_\beta.
\end{equation}
The lowest order deformation of $\hat d_0=\sum_\al \xi_\al \p_\al$ is given by
\begin{equation}
\hat d \approx \sum_{\al=1}^n \xi_\al \p_\al-\frac{1}{2}\sum_{\al,\beta,\rho=1}^n K_{\beta\al\rho}\, \xi_\rho \p_\al \p_\beta.
\end{equation}
It is important to note, however, that $\hat d$ can be expressed in the variables $\tilde \xi_\al$ and $\tilde \p_\al$ as $\hat d = \sum_{\al=1}^n \tilde \xi_\al \tilde \p_\al$
where $\tilde \xi_\al$ and $\tilde \p_\al$ are related to $\xi_\al$ and $\p_\al$ by the similarity transformations \eqref{4.87}--\eqref{4.88}.
Thus, in the new variables, the exterior derivative $\hat d$ is undeformed. Examples of differential calculi (of classical and non-classical dimension) on the
$\kappa$--deformed space and their realizations can be found in Refs. \cite{Meljanac-1,Meljanac-2,Meljanac-3,Meljanac-4,Juric-1,Juric-2}.

Finally, we note that the exterior derivative
$d=[\hat d,\, \cdot\, ]$ satisfies (i) the Leibniz rule $d(\hat f \hat g) = (d\hat f) \hat g + \hat f d\hat g$ and (ii) the nilpotency condition $d^2 \hat f=0$
for all monomials $\hat f = \hat f(\hat x)$ and $\hat g = \hat g(\hat x)$. The nilpotency of $d$ follows from the graded Jacobi identity
\begin{equation}
[\hat d, [\hat d, \hat f]]-[\hat d, [\hat f, \hat d]]+[\hat f, [\hat d, \hat d]]=0
\end{equation}
which implies $d^2 \hat f = [\hat d, [\hat d, \hat f]]=0$ since $[\hat d, \hat d]=0$. Thus, the realization \eqref{3.70-A} retains the
usual properties of the undeformed exterior derivative $\hat d_0$.

\section{Conclusion}

In this paper we have presented a general construction of a first order bicovariant differential calculus on the Lie algebra type NC space $U(\g_0)$
where $\g_0$ is a finite dimensional Lie algebra. One--forms $\theta_\al$ are introduced by extending $\g_0$ to a Lie superalgebra $\g=\g_0\oplus \g_1$
defined by Eq. \eqref{2.1} and satisfying a compatibility condition required by the exterior derivative $d\colon U(\g_0)\to \oplus_{\al=1}^n U(\g_0)\theta_\al$.
We have shown that the enveloping algebra $U(\g)$ admits an extension by shift operators $T_{AB}$ such that the commutation relations between one--forms and functions
can be expressed in terms of the action of $T_{AB}$ on $U(\g)$. In our approach the geometrical objects are constructed as deformations of the corresponding classical
notions on the Euclidean space. This is achieved by using realizations of the generators of $\g$ as formal power series in a Weyl superalgebra. We have constructed
realizations of an arbitrary Lie superalgebra \eqref{2.1}, and in the special case $\dim(\g_0)=\dim(\g_1)$ (i.e. differential calculus of classical dimension) we have also found a realization
of the exterior derivative $d$. The operator $d$ is nilpotent and it satisfies the undeformed Leibniz rule on $U(\g_0)$, as in the classical case. By using an automorphism of the
Weyl superalgebra, we conjecture that the Weyl symmetric realization \eqref{3.55} can be extended to all Lie superagebras $\g$ defined by Eq. \eqref{2.1}.

The differential algebras generated by NC coordinates $X_\mu$ and one--forms $\theta_\mu$ may have additional symmetries. It is shown in Ref. \cite{Meljanac-1}
that the differential algebras for the $\kappa$--Minkowski space are covariant under certain $\kappa$--deformations of the $\mathfrak{igl}(n)$ algebra.
The differential algebra $S_1$ defined by Eqs. \eqref{kappa} and \eqref{S1-kappa} is covariant under the Lorentz algebra $\mathfrak{so}(1,n-1)$ extended with the
dilatation operator $D$. If we denote the generators of $\mathfrak{so}(1,n-1)$ by $M_{\mu\nu}$, then
\begin{alignat}{2}
[M_{\rho\sigma},X_\nu] &= \eta_{\sigma\nu} X_\rho-\eta_{\rho\nu} X_\sigma-i(\eta_{\sigma\nu} a_\rho-\eta_{\rho\nu} a_\sigma)D, \quad &
[D,X_\nu] &= X_\nu - ia_\nu D, \\
[M_{\rho\sigma}, \theta_\nu] &=\eta_{\sigma\nu}\theta_\rho-\eta_{\rho\nu}\theta_\sigma , \quad &
[D,\theta_\nu] &= \theta_\nu,
\end{alignat}
where $\eta_{\mu\nu} = diag(-1,1,\ldots, 1)$. For a more detailed discussion of the Hopf algebra structure of the symmetry algebra generated by $M_{\mu\nu}$
and $D$, which can be obtained from twist, the reader is referred to Refs. \cite{Meljanac-1,G-2,Meljanac-8}. The method of
realizations presented here can be used to further study geometry on NC spaces from a deformation point of view. In particular, the study of field
theory on NC spaces requires introduction of higher--order forms, vector fields, the Lie and inner derivative, the Hodge $\ast$--operator and an
integral. For a general Lie algebra type NC space, these problems will be considered elsewhere.

\section*{Appendix}

\begin{lemma}\label{lm-1}
The system of differential equations
\begin{equation}\label{A-73}
\sum_{\al=1}^n \frac{\p F_{ab}}{\p \p_\al}\, \psi(\CC)_{\lambda\al} = \sum_{c=1}^m K_{a\lambda c} F_{cb}, \quad 1\leq a,b\leq m,
\end{equation}
where $F_{ab}=F_{ab}(\p)$ is a formal power series satisfying the initial condition $F_{ab}(0)=\delta_{ab}$, has a unique
solution $F_{ab}(\p)=(e^\KK)_{ab}$.
\end{lemma}

\proof Multiplying Eq. \eqref{A-73} by $\p_\lambda$ and summing over $\lambda$ we obtain
\begin{equation}
\sum_{\al=1}^n \frac{\p F_{ab}}{\p \p_\al} \Big(\sum_{\lambda=1}^n \psi(\CC)_{\lambda\al} \p_\lambda\Big) = \sum_{c=1}^m \KK_{ac} F_{cb}
\end{equation}
where $\KK_{ac}=\sum_{\lambda=1}^n K_{a\lambda c} \p_\lambda$. Expanding $\psi(\CC)$ into power series  and using antisymmetry of the structure constants,
$C_{\mu\nu\lambda}=-C_{\nu\mu\lambda}$, we find that $\sum_{\lambda=1}^n \psi(\CC)_{\lambda \al} \p_\lambda =\p_\al$. Thus, Eq. \eqref{A-73} becomes
\begin{equation}\label{A-75}
\sum_{\al=1}^n \frac{\p F_{ab}}{\p \p_\al} \p_\al = \sum_{c=1}^m \KK_{ac} F_{cb}.
\end{equation}
Consider the function $F_{ab}(t \p)$, $t\in \R$. We note that Eq. \eqref{A-75} can be written as
\begin{equation}\label{A-76}
\frac{d}{d t} F_{ab}(t \p) = \sum_{c=1}^m \KK_{ac}\, F_{cb}(t \p)
\end{equation}
when both sides of Eq. \eqref{A-76} are evaluated at $t=1$. Hence, it suffices to find a solution of Eq. \eqref{A-76} with initial condition $F_{ab}(0)=\delta_{ab}$.
We look for the solution as a power series in $t$,
\begin{equation}\label{A-77}
F_{ab}(t \p) = \sum_{k=1}^\infty F_{ab}^{(k)} (\p)\, t^k.
\end{equation}
The initial condition implies $F^{(0)}_{ab}=\delta_{ab}$. Substituting Eq. \eqref{A-77} into \eqref{A-76} and equating terms with like powers of $t$
we obtain a recurrence relation for $F_{ab}^{(k)}$,
\begin{equation}
F_{ab}^{(k)} = \frac{1}{k} \sum_{c=1}^m \KK_{ac}\, F_{cb}^{(k-1)}, \quad k\geq 1.
\end{equation}
The solution of the recurrence relation is given by
\begin{equation}
F_{ab}^{(k)} = \frac{1}{k!} (\KK^k)_{ab}, \quad k\geq 1,
\end{equation}
which yields $F_{ab}(t \p) = (e^{t \KK})_{ab}$. Hence, the solution of Eq. \eqref{A-73} with $F_{ab}(0)=\delta_{ab}$
is given by $F_{ab}(t\p)\mid_{t=1}=(e^\KK)_{ab}$. As a result, we have proved the identity
\begin{equation}
\sum_{\al=1}^n \frac{\p}{\p \p_\al}(e^\KK)_{ab}\, \psi(\CC)_{\lambda\al} = \sum_{c=1}^m K_{a\lambda c} (e^\KK)_{cb}.
\end{equation}
\qed

\begin{lemma}\label{lm-2}
Let $\phi\colon \hat \A_{(n,m)}\to \hat \A_{(n,m)}$ be the transformation defined by $\phi(u)=SuS^{-1}$ where $S=\exp\big(\sum_{a,b=1}^m \tilde \xi_a\, \tilde q_b (\ln M)_{ba}\big)$. Then the
generators of $\hat \A_{(n,m)}$ transform according to Eqs. \eqref{4.87}--\eqref{4.88}.
\end{lemma}

\proof Denote $Q=\ln M$ and let $P=\sum_{a,b=1}^m \tilde \xi_a\, \tilde q_b Q_{ba}$. For any $u\in \hat \A_{(n,m)}$ we have
\begin{equation}
\phi(u) = e^P u e^{-P} = \Big(I+\sum_{k=1}^\infty \frac{1}{k!} \ad^k(P)\Big) u.
\end{equation}
If $u=\tilde \xi_c$, one can show by induction that $\ad^k(P)\tilde \xi_c = \sum_{b=1}^m \tilde \xi_b (Q^k)_{cb}$, $k\geq 1$, which yields
\begin{equation}
\phi(\tilde \xi_c)=\sum_{b=1}^m \tilde \xi_b (e^Q)_{cb} = \sum_{b=1}^m \tilde \xi_b M_{cb}.
\end{equation}
Similarly, $\ad^k(P) \tilde q_c =(-1)^k \sum_{b=1}^m \tilde q_b (Q^k)_{bc}$, $k\geq 1$, which implies
\begin{equation}
\phi(\tilde q_c)=\sum_{b=1}^m \tilde q_b (e^{-Q})_{bc} = \sum_{b=1}^m \tilde q_b (M^{-1})_{bc}.
\end{equation}
The transformation of $\tilde \p_\mu$ is trivial since $\phi(\tilde \p_\mu)=\tilde \p_\mu$. In order to find the transformation of $\tilde x_\mu$,
one verifies by induction that
\begin{equation}
\ad^k (P) \tilde x_\mu = (-1)^k \sum_{a,b=1}^m \tilde \xi_a\, \tilde q_b \Big(\ad^{k-1}(Q)\frac{\p Q}{\p \tilde \p_\mu}\Big)_{ba}, \quad k\geq 1
\end{equation}
(where $\ad^0(P)=id$). As a result, we find that
\begin{equation}\label{A-115}
\phi(\tilde x_\mu)=\tilde x_\mu + \sum_{a,b=1}^m \tilde \xi_a\, \tilde q_b \left[\sum_{k=0}^\infty \frac{(-1)^k}{(k+1)!} \ad^k(P)\frac{\p Q}{\p \tilde \p_\mu}\right]_{ba}.
\end{equation}
The series in Eq. \eqref{A-115} can be evaluated by using F. Schur's formula
\begin{equation}
\frac{d}{dt} e^A=e^A\, \frac{I-e^{-\ad(A)}}{\ad(A)} \frac{dA}{dt}
\end{equation}
which implies
\begin{equation}\label{A-117}
e^{-A} \frac{d}{dt} e^A = \sum_{k=0}^\infty \frac{(-1)^k}{(k+1)!}\, \ad^k(A)\frac{dA}{dt}.
\end{equation}
Using Eq. \eqref{A-117} we find that expression \eqref{A-115} is explicitly given by
\begin{equation}
\phi(\tilde x_\mu) = \tilde x_\mu+\sum_{a,b=1}^m \tilde \xi_a\, \tilde q_b \Big(e^{-Q} \frac{\p e^Q}{\p \tilde \p_\mu}\Big)_{ba} = \tilde x_\mu + \sum_{a,b=1}^m \tilde \xi_a\, \tilde q_b
\Big(M^{-1} \frac{\p M}{\p \tilde \p_\mu}\Big)_{ba}.
\end{equation}
\qed

\section*{Acknowledgments}
The work of S.M. has been supported by Croatian Science Foundation under the Project No. IP--2014--09--9582 and the H2020 Twinning project No.
692194, ``RBI--T--WINNING''.

\end{document}